\documentclass[nohyperref]{article}

\usepackage{microtype}
\usepackage{graphicx}
\usepackage{subfigure}
\usepackage{booktabs} 

\usepackage{hyperref}

\usepackage[preprint]{arxiv2023}

\usepackage{amsmath}
\usepackage{amssymb}
\usepackage{mathtools}
\usepackage{amsthm}

\usepackage[capitalize,noabbrev]{cleveref}

\theoremstyle{plain}

\theoremstyle{definition}

\theoremstyle{remark}

\usepackage[utf8]{inputenc} 
\usepackage[T1]{fontenc}    
\usepackage{url}            
\usepackage{amsfonts}       
\usepackage{nicefrac}       
\usepackage{xcolor}         
\usepackage{multicol}

\usepackage{csquotes}
\usepackage{arydshln}

\newcommand{\norm}[1]{\left\lVert#1\right\rVert} 
\usepackage{caption} 
\usepackage{mathtools}
\usepackage[frozencache,cachedir=.]{minted}
\usepackage{dirtytalk}
\usepackage{enumitem}
\usepackage[textsize=tiny]{todonotes}

\icmltitlerunning{GAUCHE: A Library for Gaussian Processes in Chemistry}

\begin{document}

\twocolumn[
\icmltitle{GAUCHE: A Library for Gaussian Processes in Chemistry}



\icmlsetsymbol{equal}{*}

\begin{icmlauthorlist}
\icmlauthor{Ryan-Rhys Griffiths}{equal,cam}
\icmlauthor{Leo Klarner}{equal,ox}
\icmlauthor{Henry Moss}{equal,second}
\icmlauthor{Aditya Ravuri}{equal,cam}
\icmlauthor{Sang Truong}{equal,stanford}
\icmlauthor{Samuel Stanton}{equal,genentech}
\icmlauthor{Gary Tom}{equal,utor,vector}
\icmlauthor{Bojana Rankovic}{equal,epfl}
\icmlauthor{Yuanqi Du}{equal,cornell}
\icmlauthor{Arian Jamasb}{equal,cam}
\icmlauthor{Aryan Deshwal}{was}
\icmlauthor{Julius Schwartz}{cam}
\icmlauthor{Austin Tripp}{cam}
\icmlauthor{Gregory Kell}{kcl}
\icmlauthor{Simon Frieder}{ox}
\icmlauthor{Anthony Bourached}{ucl}
\icmlauthor{Alex J. Chan}{cam}
\icmlauthor{Jacob Moss}{cam}
\icmlauthor{Chengzhi Guo}{cam}
\icmlauthor{Johannes Durholt}{evo}
\icmlauthor{Saudamini Chaurasia}{syr}
\icmlauthor{Felix Strieth-Kalthoff}{utor}
\icmlauthor{Alpha A. Lee}{cam}
\icmlauthor{Bingqing Cheng}{ist}
\icmlauthor{Al\'{a}n Aspuru-Guzik}{utor,vector,cifar}
\icmlauthor{Philippe Schwaller}{epfl}
\icmlauthor{Jian Tang}{mila,hec,cifar}
\end{icmlauthorlist}

\icmlaffiliation{evo}{Evonik Industries AG}
\icmlaffiliation{ist}{IST Austria}
\icmlaffiliation{syr}{Syracuse University}
\icmlaffiliation{was}{Washington State University}
\icmlaffiliation{cam}{University of Cambridge}
\icmlaffiliation{ox}{University of Oxford}
\icmlaffiliation{second}{Secondmind Labs}
\icmlaffiliation{stanford}{Stanford University}
\icmlaffiliation{epfl}{EPFL}
\icmlaffiliation{cornell}{Cornell University}
\icmlaffiliation{kcl}{King's College London}
\icmlaffiliation{ucl}{University College London}
\icmlaffiliation{mila}{MILA Quebec AI Institute}
\icmlaffiliation{genentech}{Genentech} 
\icmlaffiliation{utor}{University of Toronto} 
\icmlaffiliation{vector}{Vector Institute} 
\icmlaffiliation{cifar}{CIFAR AI Research Chair}
\icmlaffiliation{hec}{HEC Montreal}

\icmlcorrespondingauthor{Ryan-Rhys Griffiths}{ryangriff123@gmail.com}

\icmlkeywords{Molecules, Chemistry, Software}

\vskip 0.3in
]



\printAffiliationsAndNotice{\icmlEqualContribution} 

\begin{abstract}
  We introduce GAUCHE, a library for GAUssian processes in CHEmistry. Gaussian processes have long been a cornerstone of probabilistic machine learning, affording particular advantages for uncertainty quantification and Bayesian optimisation. Extending Gaussian processes to chemical representations however is nontrivial, necessitating kernels defined over structured inputs such as graphs, strings and bit vectors. By defining such kernels in GAUCHE, we seek to open the door to powerful tools for uncertainty quantification and Bayesian optimisation in chemistry. Motivated by scenarios frequently encountered in experimental chemistry, we showcase applications for GAUCHE in molecular discovery and chemical reaction optimisation. The codebase is made available at \href{https://github.com/leojklarner/gauche}{https://github.com/leojklarner/gauche}
\end{abstract}

\section{Introduction}

\label{intro}

Early-stage scientific discovery is typically characterised by the 
limited availability of high-quality experimental data \cite{2018_Zhang, 2021_mrk, 2020_Thawani}, leaving lots of knowledge to discover by additional targeted experiments.
In contrast, in the big data regime, discovery offers diminishing returns as most knowledge about the space of interest has already been acquired. As such, machine learning methodologies that facilitate discovery in small data regimes such as Bayesian optimisation (BO) \cite{2018_Bombarelli, 2020_Griffiths, 2021_Shields, 2022_Du, 2023_Griffiths} and active learning (AL) \cite{2019_Zhang, 2021_Jablonka} have great potential to expedite the rate at which performant molecules, molecular materials, chemical reactions and proteins are discovered.

Currently in molecular machine learning, Bayesian neural networks (BNNs) and deep ensembles are typically used to produce uncertainty estimates for driving BO and AL loops \cite{2019_Ryu, 2019_Zhang, 2020_Hwang, 2020_Scalia}. For small datasets, however, deep neural networks are often not the model of choice \cite{tom2022calibration}. Notably, certain deep learning experts have voiced a preference for Gaussian processes (GPs) in the small data regime \cite{2011_Bengio}. Furthermore, for BO, GPs possess particularly advantageous properties; first, they admit exact as opposed to approximate Bayesian inference and second, few of their parameters need to be determined by hand. In the words of Sir David MacKay \cite{2003_MacKay}, 

\begin{displayquote}
"Gaussian processes are useful tools for automated tasks where fine tuning for each problem is not possible. We do not appear to sacrifice any performance for this simplicity.''
\end{displayquote}

The iterative model refitting required in BO makes it a prime example of such an automated task. Although BNN surrogates have been trialled for BO \cite{2015_Snoek, 2016_Springenberg}, GPs remain the model of choice as evidenced by the results of the recent NeurIPS Black-Box Optimisation Competition \cite{2021_Turner}.


Training GPs on molecular inputs, however, is non-trivial. Canonical applications of GPs assume continuous input spaces of low and fixed dimensionality. The most popular molecular input representations are SMILES/SELFIES strings \cite{1987_Anderson, 1988_Weininger, 2020_Krenn}, fingerprints \cite{2010_Rogers, probst2018probabilistic, capecchi2020one} and graphs \cite{duvenaud2015convolutional,kearnes2016molecular}. Each of these input representations poses problems for GPs. SMILES strings have variable length, fingerprints are high-dimensional and sparse bit vectors, while graphs are also a form of non-continuous input. To construct a GP framework over molecules, GAUCHE provides GPU-based implementations of kernels that operate on molecular inputs, including string, fingerprint and graph kernels. Furthermore, GAUCHE includes support for protein and chemical reaction representations and interfaces with the GPyTorch \cite{2018_Gardner} and BoTorch \cite{2020_Balandat} libraries to facilitate usage for advanced probabilistic modelling and BO.

Concretely, our contributions may be summarised as:
\begin{enumerate}
    \item We propose a GP framework for molecules and chemical reactions.
    \item We provide an open-source, GPU-enabled library building on GPyTorch \cite{2018_Gardner}, BoTorch \cite{2020_Balandat} and RDKit \cite{rdkit}.
    \item We extend the use of black box graph kernels from GraKel \cite{2020_GraKel} to GP regression via a GPyTorch interface, along with a set of graph kernels implemented in native GPyTorch to enable optimisation of the graph kernel hyperparameters under the marginal likelihood.
    \item We conduct benchmark experiments evaluating the utility of the GP framework on regression, uncertainty quantification and BO.
\end{enumerate}

    
GAUCHE includes tutorials to guide users through the tasks considered in this paper and is made available at \href{https://github.com/leojklarner/gauche}{https://github.com/leojklarner/gauche}.

\section{Background}
\label{background}


\subsection{Gaussian Processes}

\paragraph{Notation:} $\mathbf{X} \in \mathbb{R}^{n \times d}$ is a design matrix of $n$ training examples of dimension $d$. A given row $i$ of the design matrix contains a training molecule's representation $\mathbf{x}_i$. A GP is specified by a mean function, $m(\mathbf{x}) = \mathbb{E}[f(\mathbf{x})]$ and a covariance function $k(\mathbf{x}, \mathbf{x'}) = \mathbb{E}[(f(\mathbf{x}) - m(\mathbf{x}))(f(\mathbf{x'}) - m(\mathbf{x}))]$.  $K_\theta(\mathbf{X}, \mathbf{X})$ is a kernel matrix where entries are computed by the kernel function as $[K]_{ij} = k(\mathbf{x}_i, \mathbf{x}_j)$. $\theta$ represents the set of kernel hyperparameters. The GP specifies the full distribution over the function $f$ to be modelled as
\begin{align*}
f(\mathbf{x}) \sim \mathcal{GP}\big(m(\mathbf{x}), k(\mathbf{x}, \mathbf{x'})\big).
\end{align*}

\paragraph{Prediction:} At test locations $\mathbf{\mathbf{X}_*}$ the GP returns a predictive mean, $\mathbf{\bar{f_*}} = K(\mathbf{\mathbf{X}_*}, \mathbf{X})[K(\mathbf{X}, \mathbf{X}) + \sigma_{y}^2 I]^{-1} \mathbf{y}$, and a predictive uncertainty $\text{cov}(\mathbf{f_*}) = K(\mathbf{X_*}, \mathbf{X_*}) - K(\mathbf{X_*}, \mathbf{X})[K(\mathbf{X}, \mathbf{X}) + \sigma_{y}^2 I]^{-1} K(\mathbf{X}, \mathbf{X_*}).$

\paragraph{Kernels:} The choice of kernel is an important inductive bias for the properties of the function being modelled. A common choice for continuous input domains is the radial basis function kernel $$k_{\text{RBF}}(\mathbf{x}, \mathbf{x'}) = \sigma_{f}^2 \exp\bigg(\frac{-||\mathbf{x} - \mathbf{x}'||_2^{2}}{2\ell^2}\bigg),$$ where $\sigma_{f}^2$ is the signal amplitude hyperparameter (vertical lengthscale) and $\ell$ is the (horizontal) lengthscale hyperparameter. The symbol $\theta$, introduced previously, is used to represent the set of kernel hyperparameters. For molecules, bespoke kernel functions need to be defined for structured input spaces.

\paragraph{GP Training:} Hyperparameters for GPs comprise kernel hyperparameters, $\theta$, in addition to the likelihood noise, $\sigma_{y}^2$. These hyperparameters are chosen by optimising an objective function known as the negative log marginal likelihood (NLML)
\begin{align*}
\label{equation: log_lik}
\log p(\mathbf{y}| \mathbf{X}, \theta) =&  \underbrace{-\frac{1}{2} \mathbf{y}^{\top}(K_{\theta}(\mathbf{X}, \mathbf{X}) + \sigma_{y}^2I)^{-1} \mathbf{y}}_\text{encourages fit with data} \\ 
&\underbrace{-\frac{1}{2} \log | K_{\theta}(\mathbf{X}, \mathbf{X}) + \sigma_{y}^2 I |}_\text{controls model capacity} -\frac{N}{2} \log(2\pi) \nonumber.
\end{align*}
$I\sigma_{y}^2$ represents the variance of i.i.d. Gaussian noise on the observations $\mathbf{y}$. The NLML embodies Occam's razor for Bayesian model selection \cite{2001_Rasmussen} in favouring models that fit the data without being overly complex.

\subsection{Bayesian Optimisation}
In molecular discovery campaigns we are typically interested in solving problems of the form
\begin{equation*}
    \label{Eq:ProbOne}
    \mathbf{x}^{\star} = \arg\max_{\mathbf{x} \in \mathcal{X}} f(\mathbf{x}),
\end{equation*}
where $f(\cdot):\mathcal{X} \rightarrow\mathbb{R}$ is an expensive black-box function over a structured input domain $\mathcal{X}$. In our example setting the structured input domain consists of a set of molecular representations (graphs, strings, bit vectors) and the expensive black-box function is a property of interest for a given molecule that we wish to optimise. Bayesian optimisation ({BO})~\citep{1964_Kushner, 1975_Mockus, 1975_Zhilinskas, jones1998, 2010_Brochu, 2020_Grosnit} is a data-efficient methodology for determining $\mathbf{x}^{\star}$. BO operates sequentially by selecting input locations at which to query the black-box function $f$ with the aim  of identifying the optimum in as few queries as possible. Evaluations are focused into promising areas of the the space as well as areas for which we have uncertainty, a balancing act known as the exploration/exploitation trade-off.

The two components of a BO scheme are a  probabilistic surrogate model and an acquisition function. The surrogate model is typically chosen to be a GP due to its ability to maintain calibrated uncertainty estimates through exact Bayesian inference. The uncertainty estimates of the surrogate model are then leveraged by the acquisition function to propose new input locations to query. The acquisition function is a heuristic that trades off exploration and exploitation, well-known examples of which include expected improvement (EI) \cite{1975_Mockus, jones1998} and entropy search \cite{hennig2012entropy,hernandez2014predictive,wang2017max,moss2021gibbon}. After the acquisition function proposes an input location, the black-box is evaluated at that location, the surrogate model is retrained and the process repeats until a solution is obtained. Systematic reviews of the BO literature may be found in \citet{2010_Brochu, 2016_Shahriari, 2018_Frazier}.

\begin{figure*}[h]
    \centering
    \includegraphics[width=\textwidth]{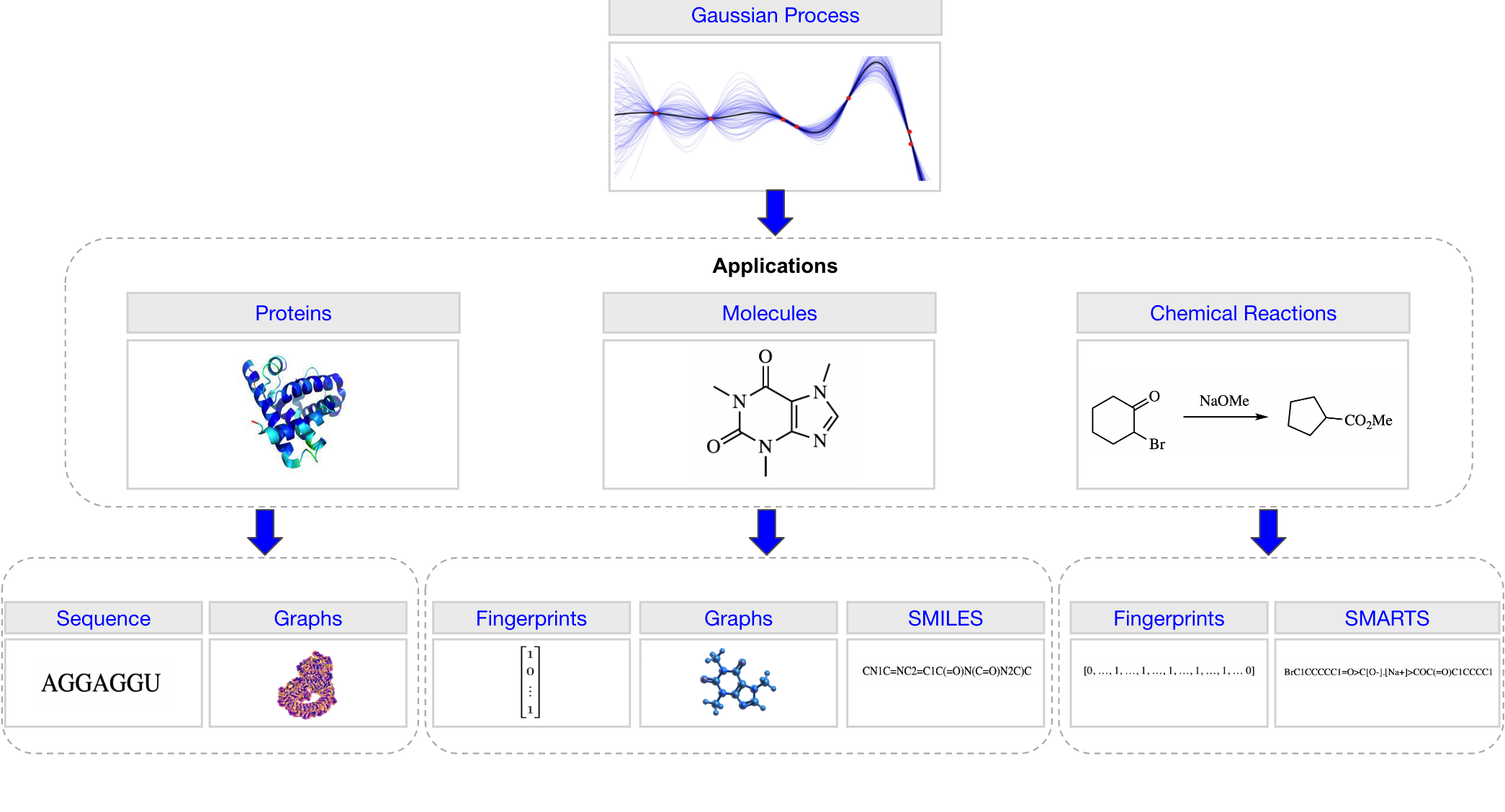}
    \caption{An overview of the applications and representations available in GAUCHE.}
    \label{overview_figure}
\end{figure*}

\subsection{Molecular Representations}

We review here the three main categories of molecular representations before describing the kernels that operate on them in section~\ref{mol_gauss}. An overview of the representations considered by GAUCHE is provided in \autoref{overview_figure}.


\paragraph{Graphs:} A molecule may be represented as an undirected, labeled graph $\mathcal{G}=(\mathcal{V}, \mathcal{E})$ where vertices $\mathcal{V}=\{v_1, \ldots, v_N\}$ represent the atoms of an $N$-atom molecule and edges $\mathcal{E}\subset\mathcal{V}\times\mathcal{V}$ represent covalent bonds between these atoms. Additional information may be incorporated in the form of vertex and edge labels $\mathcal{L}:\mathcal{V}\times\mathcal{E}\to\Sigma_V\times\Sigma_E$, with common label spaces including attributes such as atom types (e.g. hydrogen, carbon) as vertex labels and bond orders (e.g. single, double) as edge labels.

\paragraph{Fingerprints:} Molecular fingerprints were first introduced for chemical database substructure searching \cite{1993_Christie} but were later repurposed for similarity searching \cite{1990_Johnson}, clustering \cite{1997_McGregor} and classification \cite{2017_Breiman}. Extended Connectivity FingerPrints (ECFP) \cite{2010_Rogers} were introduced as part of the Pipeline project \cite{2006_Hassan} with the explicit goal of capturing features relevant for molecular property prediction \cite{2004_Xia}. ECFP fingerprints operate by assigning initial numeric identifiers to each atom in a molecule. These identifiers are subsequently updated in an iterative fashion based on the identifiers of their neighbours. The number of iterations corresponds to half the \textit{diameter} of the fingerprint and the naming convention reflects this. For example, ECFP6 fingerprints have a diameter of 6, meaning that 3 iterations of atom identifier reassignment are performed. Each level of iteration appends substructural features of increasing non-locality to an array and the array is then hashed to a bit vector reflecting the presence or absence of those substructures in the molecule.

For property prediction applications a radius of 3 or 4 is recommended. We use a radius of 3 for all experiments in the paper. Additionally we make use of fragment descriptors which are count vectors, each component of which indicates the count of a particular functional group present in a molecule. For example row 1 of the count vector could be an integer representing the number of aliphatic hydroxl groups present in the molecule. We make use of both fingerprint and fragment features computed using RDKit \cite{rdkit} as well as the concatenation of the fingerprint and fragment feature vectors, a representation termed fragprints \cite{2020_Thawani} which has shown strong empirical performance. Example representations $\mathbf{x_f}$ for fingerprints and $\mathbf{x_{fr}}$ for fragments are given as
\begin{align*}
 \mathbf{x_{f}} &= \begin{bmatrix}
       1 &
       0 & 
       \cdots &
       1
     \end{bmatrix}^\top, \:\:\:
  \mathbf{x_{fr}} = \begin{bmatrix}
       3 &
       0 &
       \cdots &
       2
     \end{bmatrix}^\top.
\end{align*}

\begin{figure}[h]
    \centering
    \includegraphics[width=0.29\textwidth]{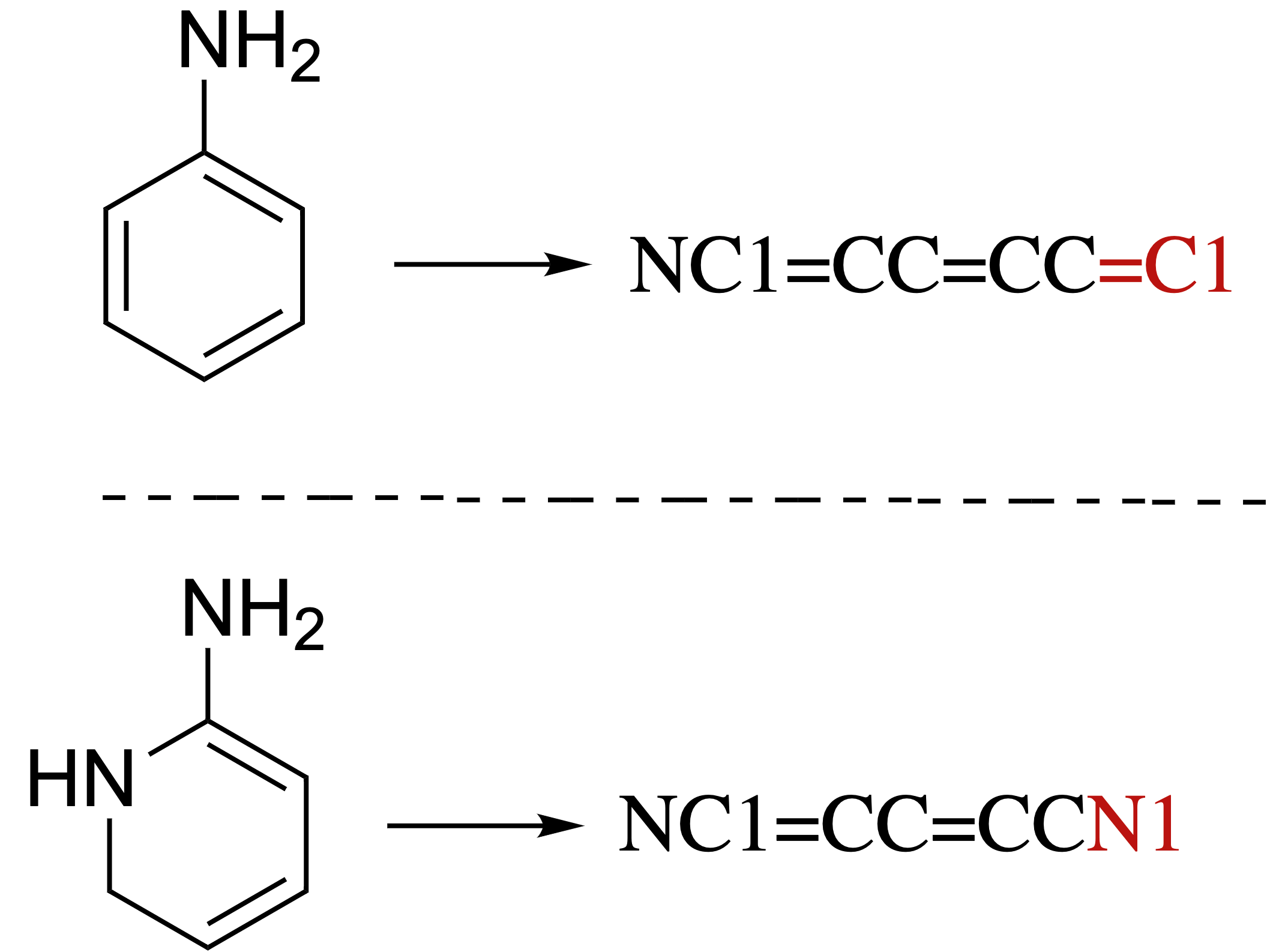}
    \caption{SMILES strings for structurally similar molecules. Similarity is encoded in the string through common contiguous subsequences (black). Local differences are highlighted in red. Molecules chosen for purposes of illustration only.}
    \label{smiles_figure}
\end{figure}

\paragraph{Strings:} The Simplified Molecular-Input Line-Entry System (SMILES) is a text-based representation of molecules \cite{1987_Anderson, 1988_Weininger}, examples of which are given in \autoref{smiles_figure}. Self-Referencing Embedded Strings (SELFIES) \cite{2020_Krenn} is an alternative string representation to SMILES such that a bijective mapping exists between a SELFIES string and a molecule.

\subsection{Reaction Representations}
Chemical reactions consist of (multiple) reactants and reagents that react to form one or more products. The choice of reactant/reagent typically constitutes a categorical design space $\mathcal{X}$. Taking as an example the high-throughput experiments by \cite{ahneman2018predicting} on Buchwald-Hartwig reactions, the reaction design space consists of 15 aryl and heteroaryl halides, 4 Buchwald ligands, 3 bases, and 23 isoxazole additives. 

\paragraph{Concatenated Molecular Representations:} If the number of choices for reactant and reagent is constant, the molecular representations discussed above may be used to encode the selected reactants and reagents, and the vectors for the individual reaction components can be concatenated to build the reaction representation \cite{ahneman2018predicting, sandfort2020structure}. An additional and commonly-used concatenated representation, is the one-hot-encoding (OHE) of the reaction categories where bits specify which of the components in the different reactant and reagent categories is present. In the Buchwald-Hartwig example, the OHE would describe which of the aryl halides, Buchwald ligands, bases and additives are used in the reaction, resulting in a 44-dimensional bit vector \cite{chuang2018comment}. 

\paragraph{Differential Reaction Fingerprints:} Inspired by the hand-engineered difference reaction fingerprints by \citet{schneider2015development}, \citet{probst2022reaction} recently introduced the differential reaction fingerprint (DRFP). This reaction fingerprint is constructed by taking the symmetric difference of the sets containing the molecular substructures on both sides of the reaction arrow. Reagents are added to the reactants. The size of the reaction bit vector generated by DRFP is independent of the number of reaction components. 

\paragraph{Data-Driven Reaction Fingerprints:} \citet{schwaller2021mapping} described data-driven reaction fingerprints using Transformer models such as BERT \cite{devlin2018bert} trained in a supervised or an unsupervised fashion on reaction SMILES. Those models can be fine-tuned on the task of interest to learn more specific reaction representations \cite{schwaller2021prediction} (RXNFP). Similar to the DRFP, the size of the data-driven reaction fingerprints is independent of the number of reaction components. 

\subsection{Protein Representations}
Proteins are large macromolecules that adopt complex 3D structures. Proteins can be represented in string form describing the underlying amino acid sequence. Graphs at varying degrees of coarseness may be used for structural representations that capture spatial and intramolecular relationships between structural elements, such as atoms, residues, secondary structures and chains. GAUCHE interfaces with Graphein \cite{2020_Jamasb}, a library for pre-processing and computing graph representations of structural biological data thereby enabling the application of graph kernel-based methods to protein structure.

\section{Molecular Kernels}
\label{mol_gauss}

Here we introduce examples of the classes of GAUCHE kernel designed to operate on the molecular representations introduced in \autoref{background}.

\subsection{Fingerprint Kernels}

\paragraph{Scalar Product Kernel:} The simplest kernel to operate on fingerprints is the scalar product or linear kernel defined for vectors $\mathbf{x}, \mathbf{x'} \in \mathbb{R}^d$ as
\begin{align*}
    k_{\text{Scalar Product}}(\mathbf{x}, \mathbf{x'}) \coloneqq \sigma_{f}^2 \cdot \langle\mathbf{x}, \mathbf{x'}\rangle,
\end{align*}
where $\sigma_{f}$ is a scalar signal variance hyperparameter and $\langle\cdot, \cdot\rangle$ is the Euclidean inner product.

\paragraph{Tanimoto Kernel:} Introduced as a general similarity metric for binary attributes \cite{1971_Gower}, the Tanimoto kernel was first used in chemoinformatics in conjunction with non-GP-based kernel methods \cite{2005_Ralaivola}. It is defined for binary vectors $\mathbf{x}, \mathbf{x'} \in \{0, 1\}^d$ for $d \geq 1$ as
\begin{align*}
    k_{\text{Tanimoto}}(\mathbf{x}, \mathbf{x'}) \coloneqq \sigma_{f}^2 \cdot \frac{\langle\mathbf{x}, \mathbf{x'}\rangle}{\norm{\mathbf{x}}^2 + \norm{\mathbf{x'}}^2 - \langle\mathbf{x}, \mathbf{x'}\rangle},
\end{align*}
where $||\cdot||$ is the Euclidean norm.

\subsection{String Kernels}
String kernels \cite{lodhi2002text, cancedda2003word} measure the similarity between strings by examining the degree at which their sub-strings differ. In GAUCHE, we implement the SMILES string kernel \cite{cao2012silico} which calculates an inner product between the occurrences of sub-strings, considering all contiguous sub-strings made from at most $n$ characters (we set $n=5$ in our experiments). Therefore, for the sub-string count featurisation $\phi : \mathcal{S} \rightarrow \mathbb{R}^p$, also known as a bag-of-characters representation \cite{jurasfky2000introduction}, the SMILES string kernel between two strings $\mathcal{S}$ and $\mathcal{S}'$ is given by
\begin{align*}
    k_{\textrm{String}}(\mathcal{S},\mathcal{S}')\coloneqq \sigma^2\cdot \langle \phi(\mathcal{S}), \phi(\mathcal{S}') \rangle.
\end{align*}
More complicated string kernels do exist in the literature, for example GAUCHE also provides an implementation of the subset string kernel \cite{2020_Boss} which allows non-contiguous matches. However, we found that the significant extra computational cost of these methods did not provide improved performance over the more simple SMILES string kernel in the context of molecular data. Note that although named the SMILES string kernel, this kernel can also be applied to any other string representation of molecules e.g. SELFIES.

\subsection{Graph Kernels}

\paragraph{Graph Kernels:} 
Graph kernel methods $\phi_\lambda:\mathcal{G}\to\mathcal{H}$ map elements from a graph domain $\mathcal{G}$ to a reproducing kernel Hilbert space (RKHS) $\mathcal{H}$, in which an inner product between a pair of graphs $g,g'\in\mathcal{G}$ is derived as a measure of similarity 
\begin{align*}
    k_{\text{Graph}}(g, g') \coloneqq \sigma^2 \cdot \langle\phi_\lambda(g),\phi_\lambda(g')\rangle_\mathcal{H},
\end{align*}
where $\lambda$ denotes kernel-specific hyperarameters and $\sigma^2$ is a scale factor.
Depending on how $\phi_\lambda$ is defined \cite{Nikolentzos_2021}, the kernel considers different substructural motifs and is characterised by different hyperparameters. Feature functions $\phi_\lambda$ from standard libraries also enable maintainable software development, as preprocessing graphs by transforming them into real-valued inputs and utilizing a linear kernel (available in most GP packages) is equivalent to the calculation above.

Frequently-employed approaches include the random walk kernel \cite{2010_Viswanathan}, given by a geometric series over the count of matching random walks of increasing length with coefficient $\lambda$,
and Weisfeiler-Lehman kernel \cite{shervashidze2011weisfeiler}, given by the inner products of label count vectors over $\lambda$ iterations of the Weisfeiler-Lehman algorithm.


\paragraph{Graph Embedding:}
Pretrained graph neural networks (GNNs)~\cite{hu2019, tom2022calibration} may also be used to embed molecular graphs in a vector space, which are reminiscent of deep kernel feature functions. Since the GNN is trained on a large amount of data, the representation it produces has the potential to be a more expressive method to encode a molecule (Note: this assumes access to a large pool of in-domain data). Given a vector representation from a pretrained GNN model, we may apply any GP kernel for continuous input spaces, such as the RBF kernel.

\section{Experiments}

We evaluate GAUCHE on regression, uncertainty quantification (UQ) and BO. The principle goal in conducting regression and UQ benchmarks is to gauge whether performance on these tasks may be used as a proxy for BO performance. BO is a powerful tool for automated scientific discovery but one would prefer to avoid model misspecification in the surrogate when deploying a scheme in the real world. We make use of the following datasets, whose labels are experimentally-determined:

\begin{itemize}
\item{The Photoswitch Dataset:} \citep{2020_Thawani}: The labels, $y$ are the values of the \textit{E} isomer $\pi-\pi^*$ transition wavelength for 392 photoswitch molecules.

\item{ESOL:} \citep{2004_Delaney}: The labels $y$ are the logarithmic aqueous solubility values for 1128 organic small molecules.

\item{FreeSolv:} \citep{2014_Mobley}: The labels $y$ are the hydration free energies for 642 molecules.

\item{Lipophilicity:} The labels $y$ are the octanol/water distribution coefficient (log D at pH 7.4) of 4200 compounds curated from the ChEMBL database \cite{2012_Gaulton, 2014_Bento}.

\item{Buchwald-Hartwig reactions:} \citep{ahneman2018predicting} The labels $y$ are the yields for 3955 Pd-catalysed Buchwald–Hartwig C–N cross-couplings. 

\item{Suzuki-Miyaura reactions:} \citep{perera2018platform} The labels $y$ are the yields for 5760 Pd-catalysed Suzuki-Miyaura C-C cross-couplings.
\end{itemize}

\subsection{Regression}
\label{subsec:regression_experiments}

The regression results for molecular property prediction are reported in Table ~\ref{table: regression} and for reaction yield prediction in Table~\ref{table: reaction} of \autoref{app_b}. The datasets are split in a train/test ratio of 80/20 (note that validation sets are not required for the GP models since hyperparameters are chosen using the marginal likelihood objective on the train set). Errorbars represent the standard error across 20 random initialisations. All GP models are trained using the L-BFGS-B optimiser \cite{1989_Liu}. If not mentioned, default settings in the GPyTorch and BoTorch libraries apply. For the SELFIES representation, some molecules could not be featurised and corresponding entries are left blank. The results of Table~\ref{table: reaction} indicate that the best choice of representation (and hence the choice of kernel) is task-dependent.

For comparison, the regression experiments are repeated for Bayesian neural networks (BNNs), and deep ensembles. The prediction and calibration results are shown in Tables \ref{table: bayes_rmse} and \ref{table: bayes_nlpd}, respectively. The BNNs are based on variational inference (VI) of the posterior distribution of the network weights \cite{blundell2015weight}, as implemented in Bayesian-Torch \cite{krishnan2022bayesiantorch}. The results are shown for a fully connected (FC)-BNN and a GNN with a final Bayesian layer \cite{2020_Hwang}. 
The deep ensemble and deep kernel GP both use 1D CNN architectures \citep{2022_Stanton}.
Further details on the deep probabilistic models are given in Appendix \ref{app_d}.

\begin{table}[h]
\caption{Molecular property prediction benchmark. RMSE values for 80/20 train/test split across 20 random trials.} \centering
\scalebox{0.55}{
\begin{tabular}{l l | c c c c}
    \toprule
    \multicolumn{2}{c|}{{\bf GP Model}} & \multicolumn{4}{c}{{\bf Dataset}}  \\
    Kernel & Representation & Photoswitch & ESOL & FreeSolv & Lipophilicity \\
    \hline
    
    Tanimoto & fragprints & ${\bf 20.9} \pm {\bf 0.7}$ & $0.71 \pm 0.01$ & $1.31 \pm 0.06 $ & $\textbf{0.67} \pm \textbf{0.01}$ \\
    & fingerprints & $23.4 \pm 0.8$ & $1.01 \pm 0.01$ & $1.93 \pm 0.09$ & $0.76 \pm 0.01$ \\ 
    & fragments & $26.3 \pm 0.8$ & $0.91 \pm 0.01$ & $1.49 \pm 0.05$ & $0.80 \pm 0.01$ \\ 
    \hdashline
    
    Scalar Product & fragprints & $22.5 \pm 0.7$ & $0.88 \pm 0.01$ & $\textbf{1.27} \pm \textbf{0.02} $ & $0.77 \pm 0.01$ \\
    & fingerprints & $24.8 \pm 0.8$ & $1.17 \pm 0.01$ & $1.93 \pm 0.07$  & $0.84 \pm 0.01$ \\
    & fragments & $36.6 \pm 1.0$ & $1.15 \pm 0.01$ &  $1.63 \pm 0.03$ & $0.97. \pm 0.01$ \\
    \hdashline
    
    String & SELFIES & $24.9 \pm 0.6$ & - & - & - \\
    & SMILES & $24.8 \pm 0.7$ & $\textbf{0.66} \pm \textbf{0.01}$  & $1.31 \pm 0.01$ & $\textbf{0.68} \pm \textbf{0.01}$  \\
    
    \hdashline
    WL Kernel (GraKel) & graph & $22.4 \pm 1.4$ & $1.04 \pm 0.02$ & $1.47 \pm 0.06$ & $0.74 \pm 0.05$ \\
    
    
    
    \bottomrule
\end{tabular}
}
\label{table: regression}
\end{table}

\subsection{Uncertainty Quantification (UQ)}

To quantify the quality of the uncertainty estimates we use three metrics, the negative log predictive density (NLPD), the mean standardised log loss (MSLL) and the quantile coverage error (QCE). We provide the NLPD results in Table~\ref{table: nlpd} and defer the MSLL and QCE results to \autoref{app_a}. One trend to note is that uncertainty estimate quality is roughly correlated with regression performance.

\begin{table}[h]
\caption{UQ benchmark. NLPD values for 80/20 train/test split across 20 random trials.}
\centering
\scalebox{0.55}{
\begin{tabular}{l l | c c c c}
    \toprule
    \multicolumn{2}{c|}{{\bf GP Model}} & \multicolumn{4}{c}{{\bf Dataset}}  \\
    Kernel & Representation & Photoswitch & ESOL & FreeSolv & Lipophilicity \\
    \hline

    Tanimoto & fragprints & $\textbf{0.22} \pm \textbf{0.03}$ & $0.33 \pm 0.01$ & $0.28 \pm 0.02$ & $\textbf{0.71} \pm \textbf{0.01}$ \\
    & fingerprints & $0.33 \pm 0.03$ & $0.71 \pm 0.01$ & $0.58 \pm 0.03$ & $0.85 \pm 0.01$ \\ 
    & fragments & $0.50 \pm 0.04$ & $0.57 \pm 0.01$ & $0.44 \pm 0.03$ & $0.94 \pm 0.02$\\ 
    \hdashline
    
    Scalar Product & fragprints & $\textbf{0.23} \pm \textbf{0.03}$ & $0.53 \pm 0.01$ & $0.25 \pm 0.02$ & $0.92 \pm 0.01$ \\
    & fingerprints & $0.33 \pm 0.03$ & $0.84 \pm 0.01$ & $0.64 \pm 0.03$ & $1.03 \pm 0.01$\\
    & fragments & $0.80 \pm 0.03$ & $0.82 \pm 0.01$ & $0.54 \pm 0.02$ & $0.88 \pm 0.10$  \\
    \hdashline
    
    String & SELFIES & $0.37 \pm 0.04$ & - & - & - \\
    & SMILES & $0.30 \pm 0.04$ & $\textbf{0.29} \pm \textbf{0.03}$  & $\textbf{0.16} \pm \textbf{0.02}$  & $\textbf{0.72} \pm \textbf{0.01}$ \\
    \hdashline
    
    WL Kernel (GraKel) & graph & $0.39 \pm 0.11$ & $0.76 \pm 0.001$ & $0.47 \pm 0.02$ & - \\
    
    \bottomrule
    
\end{tabular}
}
\label{table: nlpd}
\end{table}

\begin{table}[h]
\caption{Molecular property prediction regression benchmark for deep Bayesian models. RMSE of predictions for Bayesian Neural Networks (BNN) and Bayesian Graph Neural Networks (GNN) for 80/20 train/test split across 20 random trials.}
\centering
\scalebox{0.58}{
\begin{tabular}{l l | c c c c}
    \toprule
    \multicolumn{2}{c|}{{\bf  Deep Probabilistic Models }} & \multicolumn{4}{c}{{\bf Dataset}}  \\
    Model & Representation & Photoswitch & ESOL & FreeSolv & Lipophilicity \\
    \hline

    FC-BNN & fragprints & $\textbf{20.9} \pm \textbf{0.6}$ & $0.88 \pm 0.01$ & $1.39 \pm 0.03$ & $0.75 \pm 0.01$ \\
    & fingerprints & $22.4 \pm 0.7$ & $1.08 \pm 0.02$ & $1.93 \pm 0.07$ & $0.81 \pm 0.01$ \\ 
    & fragments & $25.8 \pm 0.7$ & $1.03 \pm 0.01$ & $1.48 \pm 0.02 $ & $0.87 \pm 0.01$\\ 
    \hdashline
    
    GNN-BNN & graph & $28.5 \pm 1.2$ & $0.88 \pm 0.01$ & $\textbf{0.96} \pm \textbf{0.01}$ & $\textbf{0.73} \pm \textbf{0.02}$ \\

    \hdashline
    CNN Ensemble & SELFIES & $26.4 \pm 1.0$ & $\textbf{0.67} \pm \textbf{0.01}$ & $1.29 \pm 0.04$ & $\textbf{0.75} \pm \textbf{0.01}$ \\

        \hdashline
    CNN DKL GP & SELFIES & $25.1 \pm 0.8$ & $0.94 \pm 0.04$ & $1.41 \pm 0.11$ & $0.91 \pm 0.01$ \\
    
    \bottomrule
\end{tabular}
}
\label{table: bayes_rmse}
\end{table}

\begin{table}[h]
\caption{UQ benchmark. NLPD values of predictions for Bayesian Neural Networks (BNN) and Bayesian Graph Neural Networks (GNN) for 80/20 train/test split across 20 random trials.}
\centering
\scalebox{0.58}{
\begin{tabular}{l l | c c c c}
    \toprule
    \multicolumn{2}{c|}{{\bf  Deep Probabilistic Models }} & \multicolumn{4}{c}{{\bf Dataset}}  \\
    Model & Representation & Photoswitch & ESOL & FreeSolv & Lipophilicity \\
    \hline

    FC-BNN & fragprints & $1.63 \pm 0.44$ & $1.70 \pm 0.11$ & $1.41 \pm 0.38$ & $3.82 \pm 0.12$ \\
    & fingerprints & $2.22 \pm 0.56$ & $2.59 \pm 0.40$ & $2.65 \pm 0.72$ & $3.74 \pm 0.10$ \\
    & fragments & $0.69 \pm 0.09$ & $1.93 \pm 0.28$ & $0.89 \pm 0.12$ & $5.52 \pm 0.23$\\ 
    \hdashline
    
    GNN-BNN & graph & $1.00 \pm 0.13$ & $1.70 \pm 0.11$ & $1.01 \pm 0.02$ & $\textbf{1.14} \pm \textbf{0.01}$ \\

    \hdashline
    CNN Ensemble & SELFIES & $4.34 \pm 0.55$ & $2.91 \pm 0.14$ & $2.24 \pm 0.21$ & $2.60 \pm 0.06$ \\

    \hdashline
    CNN DKL GP & SELFIES & $\textbf{0.48} \pm \textbf{0.05}$ & $\textbf{0.90} \pm \textbf{0.15}$ & $\textbf{0.33} \pm \textbf{0.04}$ & $1.46 \pm 0.03$ \\

    \bottomrule
\end{tabular}
}
\label{table: bayes_nlpd}
\end{table}


\subsection{Bayesian Optimisation}

\begin{figure*}[]
\centering
\subfigure[Photoswitch]{\label{fig:4}\includegraphics[width=0.33\textwidth]{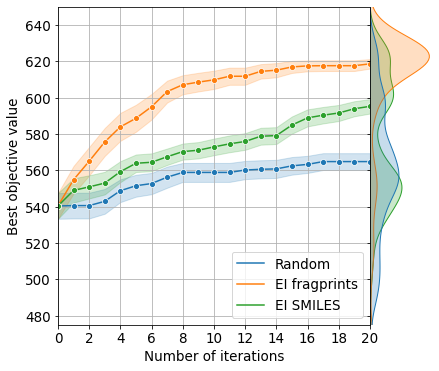}}
\subfigure[ESOL]{\label{fig:3}\includegraphics[width=0.33\textwidth]{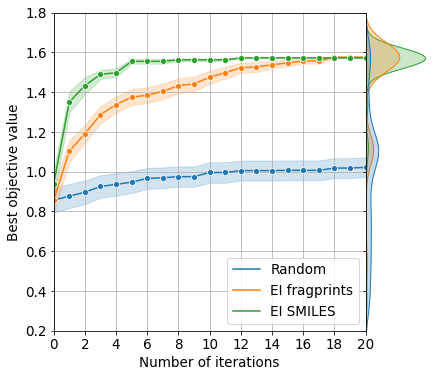}}
\subfigure[Buchwald-Hartwig reactions]{\label{fig:5}\includegraphics[width=0.33\textwidth]{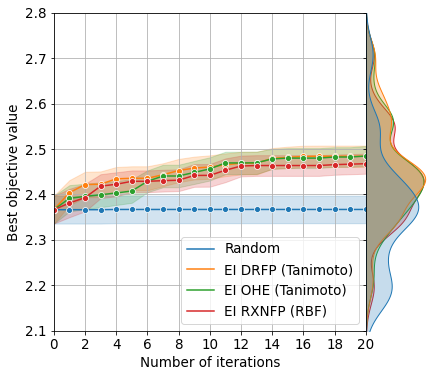}}
\caption{BO performance reporting the standard error from 50 randomly initialised trials (20 for Buchwald-Hartwig). A kernel density estimate over the trials is shown on the right axis. EI fragprints results use the Tanimoto kernel.}
\label{bayesopt}
\end{figure*}

We take forward two of the best-performing kernels, the Tanimoto-fragprint kernel and the SMILES string kernel to undertake BO over the photoswitch and ESOL datasets. Random search is used as a baseline. BO is run for 20 iterations of sequential candidate selection (EI acquisition) where candidates are drawn from 95\% of the dataset. The models are initialised with 5\% of the dataset. In the case of the photoswitch dataset this corresponds to just 19 molecules. The results are provided in \autoref{bayesopt}. In this ultra-low data setting, common to many areas of synthetic chemistry \cite{2020_Thawani} both models outperform random search, highlighting the real-world use-case for such models in supporting human chemists prioritise candidates for synthesis. Furthermore, one may observe that BO performance is tightly coupled to regression and UQ performance. In the case of the photoswitch dataset, the better-performing Tanimoto model on regression and UQ also achieves better BO performance. Additionally, we run BO on the Buchwald-Hartwig dataset using the Tanimoto kernel for the bit-vector representations DRFP and OHE, and the RBF kernel for RXNFP. All three representations perform similarly, and outperform the random search.

\section{Related Work}
General-purpose GP and BO libraries do not cater for molecular representations. Likewise, general-purpose molecular machine learning libraries do not consider GPs and BO. Here, we review existing libraries, highlighting the niche GAUCHE fills in bridging the GP and molecular machine learning communities. 

The closest work to ours is FlowMO \cite{2020_Moss}, which introduces a basic molecular GP library in the GPflow framework. In this project, we extend the scope of the library to a broader class of molecular representations (graphs), problem settings (BO) and applications (reaction optimisation and protein engineering).

\paragraph{Gaussian Process Libraries:} GP libraries include GPy (Python) \cite{2014_gpy}, GPflow (TensorFlow) \cite{2017_Matthews, 2020_Wilk}, GPyTorch (PyTorch) \cite{2018_Gardner} and GPJax (Jax) \cite{pinder2022gpjax} while examples of recent BO libraries include BoTorch (PyTorch) \cite{2020_Balandat}, Dragonfly (Python) \cite{2020_Kandasamy}, HEBO (PyTorch) \cite{2020_Rivers} and Trieste (Tensorflow) \cite{Berkeley_Trieste_2022}. The aforementioned libraries do not explicitly support molecular representations. Extension to cover molecular representations however is nontrivial, requiring implementations of bespoke GP kernels for bit vector, string and graph inputs together with modifications to BO schemes to consider acquisition function evaluations over a discrete set of heldout molecules, a setting commonly encountered in virtual screening \cite{2020_Pyzer, 2022_Graff}.

\paragraph{Molecular Machine Learning Libraries:} Molecular machine learning libraries include DeepChem \cite{2019_Ramsundar}, DGL-LifeSci \cite{2021_Li} and TorchDrug \cite{2022_Zhu}. DeepChem features a broad range of model implementations and tasks, while DGL-LifeSci focuses on graph neural networks. TorchDrug caters for applications including property prediction, representation learning, retrosynthesis, biomedical knowledge graph reasoning and molecule generation.

GP implementations are not included, however, in the aforementioned libraries. In terms of atomistic systems, DScribe \cite{2020_Himanen} features, amongst other methods, the Smooth Overlap of Atomic Positions (SOAP) representation \cite{2013_Bartok} which is typically used in conjunction with a GP model to learn atomistic properties. Automatic Selection And Prediction (ASAP) \cite{2020_Cheng} also principally focusses on atomistic properties as well as dimensionality reduction and visualisation techniques for materials and molecules. Lastly, the Graphein library focusses on graph representations of proteins \cite{2020_Jamasb}.

\paragraph{Graph Kernel Libraries:} Graph kernel libraries include GraKel \cite{2020_GraKel}, graphkit-learn \cite{2021_Linlin}, graphkernels \cite{2018_Sugiyama}, graph-kernels \cite{2015_Sugiyama}, pykernels (\href{https://github.com/gmum/pykernels}{https://github.com/gmum/pykernels}) and ChemoKernel \cite{2012_Gauzere}. The aforementioned libraries focus on CPU implementations in Python. Extending graph kernel computation to GPUs has been noted as an important direction for future research \cite{2018_Ghosh}. In our work, we build on the GraKel library by interfacing it with GPyTorch, facilitating GP regression with GPU computation. Furthermore, we enable the graph kernel hyperparameters to be learned through the marginal likelihood objective as opposed to being pre-specified and fixed upfront. It is worth noting that GAUCHE extends the applicability of GPU-enabled GPs to general graph-structured inputs beyond just molecules and proteins.

\paragraph{Molecular Bayesian Optimisation:} BO over molecular space can be divided into two classes. In the first class, molecules are encoded into the latent space of a variational autoencoder (VAE) \cite{2018_Bombarelli}. BO is then performed over the continuous latent space and queried molecules are decoded back to the original space. Much work on VAE-BO has focussed on improving the synergy between the surrogate model and the VAE \cite{2018_Griffiths, 2020_Griffiths, 2020_Tripp, 2021_Deshwal, 2021_Grosnit, 2021_Verma, 2022_Maus, 2022_Stanton}. One of the defining characteristics of VAE-BO is that it enables the generation of new molecular structures.\\

In the second class of methods, BO is performed directly over the original discrete space of molecules. In this setting it is not possible to generate new structures and so a candidate set of queryable molecules is defined. The inability to generate new structures however, is not a bottleneck to molecule discovery as the principle concern is how best to explore existing candidate sets. These candidate sets are also known as molecular libraries in the virtual screening literature \cite{2015_Pyzer}. 

To date, there has been little work on BO directly over discrete molecular spaces. In \citet{2020_Boss}, the authors use a string kernel GP trained on SMILES to perform BO to select from a candidate set of molecules. In \citet{2020_Korovina}, an optimal transport kernel GP is used for BO over molecular graphs. In \citet{2021_Hase} a surrogate based on the Nadarya-Watson estimator is defined such that the kernel density estimates are inferred using BNNs. The model is then trained on molecular descriptors. Lastly, in \citet{2017_Lobato} and \citet{2021_Moss} a BNN and a sparse GP respectively are trained on fingerprint representations of molecules. In the case of the sparse GP the authors select an ArcCosine kernel. It is a longstanding aim of the GAUCHE Project to compare the efficacy of VAE-BO against vanilla BO on real-world molecule discovery tasks.


\paragraph{Chemical Reaction Optimisation:} Chemical reactions describe how reactants transform into products. Reagents (catalysts, solvents, and additives) and reaction conditions heavily impact the outcome of chemical reactions. Typically the objective is to maximise the reaction yield (the amount of product compared to the theoretical maximum) \cite{ahneman2018predicting}, in asymmetric synthesis, where the reactions could result in different enantiomers, to maximise the enantiomeric excess \cite{zahrt2019prediction}, or to minimise the E-factor, which is the ratio between waste materials and the desired product \cite{schweidtmann2018machine}. 

A diverse set of studies have evaluated the optimisation of chemical reactions in single and  multi-objective settings \cite{schweidtmann2018machine,muller2022automated}. \citet{felton2021summit} and \citet{hase2021olympus} benchmarked reaction optimisation algorithms in low-dimensional settings including reaction conditions, such as time, temperature, and concentrations. \citet{2021_Shields} suggested BO as a general tool for chemical reaction optimisation and benchmarked their approach against human experts. \citet{haywood2021kernel} compared the yield prediction performance of different kernels and \citet{pomberger2022} the impact of various molecular representations. 

In all reaction optimisation studies above, the representations of the different categories of reactants and reagents are concatenated to generate the reaction input vector, which could lead to limitations if another type of reagent is suddenly considered. Moreover, most studies concluded that simple one-hot encodings (OHE) perform at least on par with more elaborate molecular representations in the low-data regime \cite{2021_Shields, pomberger2022, hickman2022}. In GAUCHE, we introduce reaction fingerprint kernels, based on existing reaction fingerprints \cite{schwaller2021mapping, probst2022reaction} and work independently of the number of reactant and reagent categories. 

\newpage

\section{Conclusions and Future Work}

We have introduced GAUCHE, a library for GAUssian Processes in CHEmistry with the aim of providing tools for uncertainty quantification and Bayesian optimisation that may hopefully be deployed for screening in laboratory settings. Future work is centred along two axes:

\begin{enumerate}
\item \textbf{Methodology}: We seek to incorporate more sophisticated GP-based optimisation and active learning loops in chemistry applications \cite{eyke2020iterative, 2022_Rankovic, 2023_Griffiths}, such as the application of ideas from batch \cite{gonzalez2016batch}, multi-task \cite{swersky2013multi}, multi-fidelity \cite{moss2020mumbo}, multi-objective \cite{knowles2006parego}, and quantile \cite{torossian2020bayesian} optimisation. We will also investigate the use of approximate Gaussian processes \cite{titsias2009variational, hensman2013gaussian} and their recent adaptations for BO \cite{maddox2021conditioning,chang2022fantasizing, moss2022information, moss2023IPA} to allow our methods to handle larger evaluation budgets.
\item \textbf{User Feedback}: We seek to further grow our userbase and solicit feedback from laboratory practitioners on the most common use-cases for BO and GP modelling in molecular discovery campaigns. Of particular interest is the relative prevalance of molecule generation and optimisation via VAE-BO compared to AL/BO to prioritise molecules in high-throughput screening settings.
\end{enumerate}




\newpage

\bibliography{example_paper}
\bibliographystyle{icml2023}

\newpage
\appendix

\newpage

\renewcommand{\thetable}{\Alph{section}\arabic{table}}
\setcounter{table}{0}
\onecolumn


\centering\includegraphics[height=10\baselineskip]{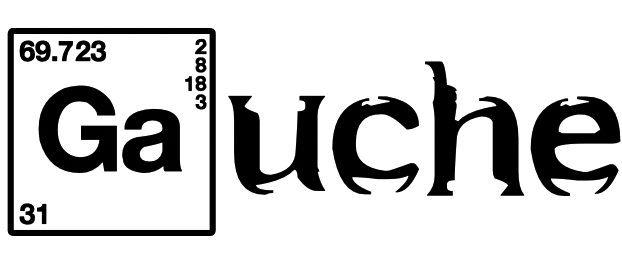}

\begin{enumerate}[label=\arabic*,leftmargin=*,labelsep=2ex,ref=\arabic*]
\item Introduction \dotfill 1
\item Background \dotfill 2
  \begin{enumerate}[label*=.\arabic*,leftmargin=*,labelsep=2ex]
    \item Gaussian Processes \dotfill 2
    \begin{enumerate}[label*=.\arabic*,leftmargin=*,labelsep=2ex]
        \item Notation
        \item Prediction
        \item Kernels
        \item GP Training
    \end{enumerate}
    \item Bayesian Optimisation \dotfill 2
    \item Molecular Representations \dotfill 3
    \begin{enumerate}[label*=.\arabic*,leftmargin=*,labelsep=2ex]
        \item Graphs
        \item Fingerprints
        \item Strings
    \end{enumerate}
    \item Reaction Representations \dotfill 3
    \begin{enumerate}[label*=.\arabic*,leftmargin=*,labelsep=2ex]
        \item Concatenated Molecular Representations
        \item Differential Reaction Fingerprints
    \end{enumerate}
    \item Protein Representations \dotfill 4
  \end{enumerate}
\item Molecular Kernels \dotfill 4
    \begin{enumerate}[label*=.\arabic*,leftmargin=*,labelsep=2ex]
        \item Fingerprint Kernels
        \begin{enumerate}[label*=.\arabic*,leftmargin=*,labelsep=2ex]
        \item Scalar Product Kernel
        \item Tanimoto Kernel
        \end{enumerate}
        \item String Kernels
        \item Graph Kernels
        \begin{enumerate}[label*=.\arabic*,leftmargin=*,labelsep=2ex]
        \item Graph Kernel
        \item Graph Embedding
        \end{enumerate}
    \end{enumerate}
\item Experiments \dotfill 5
    \begin{enumerate}[label*=.\arabic*,leftmargin=*,labelsep=2ex]
        \item Regression
        \item Uncertainty Quantification (UQ)
        \item Bayesian Optimisation
    \end{enumerate}
\item Related Work \dotfill 6
    \begin{enumerate}[label*=.\arabic*,leftmargin=*,labelsep=2ex]
        \item Gaussian Process Libraries
        \item Molecular Machine Learning Libraries
        \item Graph Kernel Libraries
        \item Molecular Bayesian Optimisation
        \item Chemical Reaction Optimisation
    \end{enumerate}
\item Conclusions and Future Work \dotfill 8
\end{enumerate}

\begin{enumerate}[label=\Alph*,leftmargin=*,labelsep=2ex]
\item Coding Kernels in GAUCHE \dotfill 16
\item Chemical Reaction Yield Prediction Experiments \dotfill 16
\item Uncertainty Quantification Experiments \dotfill 16
\item Deep Probabilistic Models \dotfill 17
\end{enumerate}

\newpage

\section{Coding Kernels in GAUCHE}
\label{app:a section}

We provide an example of the class definition for the Tanimoto kernel in GAUCHE below

\begin{minted}{python}
class TanimotoGP(ExactGP):
    def __init__(self, train_x, train_y, likelihood):
        super(TanimotoGP, self).__init__(train_x,
                                         train_y,
                                         likelihood)
        self.mean_module = ConstantMean()
        # We use the Tanimoto kernel to work with
        # molecular fingerprint representations
        self.covar_module = ScaleKernel(TanimotoKernel())

    def forward(self, x):
        mean_x = self.mean_module(x)
        covar_x = self.covar_module(x)
        return MultivariateNormal(mean_x, covar_x)
\end{minted}

and an example definition of a black box kernel (where gradients with respect to hyperparameters and input labels are not required).

\begin{minted}{python}

class WLKernel(gauche.Kernel):
    def __init__(self):
        super().__init__()
        self.kernel = grakel.kernels.WeisfeilerLehman()

    @lru_cache(maxsize=3)
    def kern(self, X):
        return tensor(self.kernel.fit_transform(X.data))

class GraphGP(gauche.SIGP):
    def __init__(self, train_x, train_y, likelihood):
        super().__init__(train_x, train_y, likelihood)
        self.mean = ConstantMean()
        self.covariance = WLKernel()

    def forward(self, X):
        # X is a gauche.Inputs instance, with X.data
        # holding a list of grakel.Graph instances.
        mean = self.mean(zeros(len(X.data), 1))
        covariance = self.covariance(X)
        return MultivariateNormal(mean, covariance)
\end{minted}

Importantly, GAUCHE inherits all the facilities of GPyTorch and GraKel allowing a broad range of of models to be defined on molecular inputs such as deep GPs, multioutput GPs and heteroscedastic GPs.

\section{Chemical Reaction Yield Prediction Experiments}
\label{app_b}

Further regression and uncertainty quantification experiments are presented in Table~\ref{table: reaction}. The differential reaction fingerprint in conjunction with the Tanimoto kernel is the best-performing reaction representation.

\begin{table*}[ht]
\caption{Chemical reaction regression benchmark. 80/20 train/test split across 20 random trials.}
\centering
\begin{tabular}{l l | c c c c}
    \toprule
    \multicolumn{2}{c|}{\bf GP Model} & \multicolumn{4}{c}{\bf Buchwald-Hartwig}  \\
    Kernel & Representation & RMSE $\downarrow$ & $R^{2}$ score $\uparrow$ & MSLL $\downarrow$ & QCE $\downarrow$ \\ \hline
    
    Tanimoto & OHE & $7.94 \pm 0.05$ & $0.91 \pm 0.001$ & $-0.06 \pm 0.002$& $0.011 \pm 0.001$\\
    & DRFP & $\textbf{6.48} \pm \textbf{0.45}$ & $\textbf{0.94} \pm \textbf{0.015}$ & $\textbf{-0.15} \pm \textbf{0.07} $& $0.027 \pm 0.002$ \\ 
    \hdashline
    
    Scalar Product & OHE & $15.23 \pm 0.052$ & $0.69 \pm 0.002$ & $0.57 \pm 0.002$ & $0.008 \pm 0.001$\\ 
    & DRFP & $14.63 \pm 0.050$ & $0.71 \pm 0.002$ &$0.55 \pm 0.002$ & $0.010 \pm 0.001$ \\
    \hdashline
    
    RBF & RXNFP & $10.79 \pm 0.049$ &$0.84 \pm 0.001 $ & $0.37 \pm 0.005$ & $0.024 \pm 0.001$ \\
    \hline
    
    \multicolumn{2}{c|}{} & \multicolumn{4}{c}{\bf Suzuki-Miyaura}  \\ \hline
    
    Tanimoto & OHE & $11.18 \pm 0.036$ & $0.83 \pm 0.001$ & $0.23 \pm 0.001 $ & $0.007 \pm 0.001 $  \\
    & DRFP & $11.46 \pm 0.038 $ & $0.83 \pm 0.001 $ &$0.25 \pm 0.006$ &$0.019 \pm 0.000$  \\
    \hdashline
    
    Scalar Product & OHE & $19.91 \pm 0.042$ & $0.47 \pm 0.003$ & $0.82 \pm 0.001 $  & $0.012 \pm 0.001$ \\
    & DRFP & $19.66 \pm 0.042$ & $0.52 \pm 0.003$ & $0.81 \pm 0.001$  & $0.014 \pm 0.001$  \\
    \hdashline
    
    RBF & RXNFP & $13.83 \pm 0.048 $ & $0.75 \pm 0.002$ & $0.50 \pm 0.001$ & $0.007 \pm 0.001$ \\
    \bottomrule
\end{tabular}
\label{table: reaction}
\end{table*}

\section{Uncertainty Quantification Experiments}
\label{app_a}

In Table~\ref{table: MSLL} and Table~\ref{table: QCE} we present further uncertainty quantification metrics. Numerical errors were encountered with the WL kernel on the large lipophilicity dataset which invalidated the results and so the corresponding entry is left blank. The native random walk kernel was discontinued (for the time being) due to poor performance!

\begin{table}[h]
\caption{UQ Benchmark. MSLL Values ($\downarrow$) for 80/20 Train/Test Split.}
\centering
\begin{tabular}{l l | c c c c}
    \toprule
    \multicolumn{2}{c|}{{\bf GP Model}} & \multicolumn{4}{c}{{\bf Dataset}}  \\
    Kernel & Representation & Photoswitch & ESOL & FreeSolv & Lipophilicity \\
    \hline
    
    Tanimoto & fragprints & $\textbf{0.06} \pm \textbf{0.01}$ & $0.17 \pm 0.04$ & $0.16 \pm 0.02$ & $\textbf{0.50} \pm \textbf{0.006}$ \\
    & fingerprints & $0.16 \pm 0.01$ & $0.55 \pm 0.01$ & $0.42 \pm 0.02$ & $0.63 \pm 0.004$ \\ 
    & fragments & $0.27 \pm 0.01$ & $0.34 \pm 0.04$ & $0.24 \pm 0.02$& $0.72 \pm 0.003$\\ 
    \hdashline
    
    Scalar Product & fragprints & $0.03 \pm 0.01$ & $0.32 \pm 0.004$ & $0.06 \pm 0.01$ & $0.67 \pm 0.003$ \\
    & fingerprints & $0.11 \pm 0.01$ & $0.64 \pm 0.006$ & $0.41 \pm 0.02$ & $0.79 \pm 0.003$\\
    & fragments & $0.56 \pm 0.01$ & $0.58 \pm 0.005$ & $0.29 \pm 0.01$ & $0.94 \pm 0.003$  \\
    \hdashline
    
    String & SELFIES & $0.13 \pm 0.01$ & - & - & - \\
    & SMILES & $\textbf{0.08} \pm \textbf{0.02}$ & $\textbf{0.03} \pm \textbf{0.005}$  & $\textbf{0.03} \pm \textbf{0.02}$  & $\textbf{0.52} \pm \textbf{0.002}$ \\
    \hdashline
    
    WL Kernel (GraKel) & graph & $0.14 \pm 0.03$ & $0.54 \pm 0.01$ & $0.26 \pm 0.01$ & - \\

    \bottomrule
    
\end{tabular}
\label{table: MSLL}
\end{table}

\begin{table}[h]
\caption{UQ benchmark. QCE values ($\downarrow$) for 80/20 train/test split across 20 random trials.}
\centering
\begin{tabular}{l l | c c c c}
    \toprule
    \multicolumn{2}{c|}{{\bf GP Model}} & \multicolumn{4}{c}{{\bf Dataset}}  \\
    Kernel & Representation & Photoswitch & ESOL & FreeSolv & Lipophilicity \\
    \hline
    
    Tanimoto & fragprints & $\textbf{0.019} \pm \textbf{0.003}$ & $0.023 \pm 0.002$ & $0.023 \pm 0.002$ & $0.006 \pm 0.002$ \\
    & fingerprints & $0.023 \pm 0.003$ & $0.022 \pm 0.002$ & $0.018 \pm 0.003$ & $0.006 \pm 0.001$ \\ 
    & fragments & $0.025 \pm 0.005$ & $0.012 \pm 0.002$ & $0.014 \pm 0.002$& $0.009 \pm 0.002$\\ 
    \hdashline
    
    Scalar Product & fragprints & $0.033 \pm 0.006$ & $0.010 \pm 0.002$ & $0.017 \pm 0.003$ & $0.010 \pm 0.001$ \\
    & fingerprints & $0.036 \pm 0.006$ & $0.014 \pm 0.002$& $0.016 \pm 0.002$ & $0.009 \pm 0.001$ \\
    & fragments & $0.027 \pm 0.004$ & $0.012 \pm 0.003$ & $0.021 \pm 0.003$ & $0.010 \pm 0.001$ \\
    \hdashline
    
    String & SELFIES & $0.031 \pm 0.006$ & -  & - & - \\
    & SMILES & $0.024 \pm 0.003$ & $0.016 \pm 0.002$  & $0.019 \pm 0.003$  & $0.005 \pm 0.001$ \\
    \hdashline
    
    WL Kernel (GraKel) & graph & $0.025 \pm 0.007$ & $0.011 \pm 0.004$ & $0.019 \pm 0.009$ & $0.066 \pm 0.014$ \\
    \bottomrule
\end{tabular}
\label{table: QCE}
\end{table}

\newpage

\section{Deep Probabilistic Models}

\label{app_d}

There is a growing body of work applying deep learning to molecular property prediction \citep{walters2020applications}.
Therefore in addition to evaluating GPs with varying shallow kernel functions, we repeat the regression experiments with a range of deep Bayesian models, varying both the network architecture and the Bayesian inference procedure. 
We evaluate the following models:
\begin{itemize}
    \item \textbf{FC-BNN + VI} is a fully connected neural network with a single variational inference (VI) Bayesian layer with 100 nodes, followed by the rectified linear unit (ReLU) activation, and a final output layer. Trained with early stopping.
    \item \textbf{GNN-BNN + VI} utilises the same network and graph features as used for the graph embeddings \cite{hu2019} followed by a final VI Bayesian layer. Trained with early stopping.
    \item \textbf{CNN DKL GP} is the same approach and architecture used by \citet{2022_Stanton} to predict molecular properties for Bayesian optimisation. Using SELFIES representations, a 1D CNN encoder is shared and trained jointly through a generative masked language model (MLM) head \citep{devlin2018bert} and a discriminative deep kernel GP head \citep{wilson2016deep}.
    \item \textbf{CNN Ensemble} is a deep ensemble of 1D CNN networks, also implemented by \citet{2022_Stanton}, where each ensemble component uses the SELFIES molecule representation and is trained independently to minimize the MSE loss. Deep ensembles have been shown to provide high-fidelity approximations of Bayesian model averages relative to alternative approaches such as Laplace approximation or VI \citep{izmailov2021bayesian}.
\end{itemize}

To summarize our results in Tables \ref{table: bayes_rmse} \& \ref{table: bayes_nlpd}, we find that deep Bayesian models generally do not outperform discrete string kernels or shallow continuous kernels on hand-crafted features (e.g. fragprints) on the datasets we consider, although in some cases performance is very competitive.
Our results suggest that for small to mid-sized molecular datasets the Tanimoto kernel combined with fragprint representations in particular is a very compelling option, with good accuracy, calibration, and runtime across all tasks.


\end{document}